\documentclass[twocolumn,english,aps.two coloumn]{revtex4-1}
\usepackage{ae,aecompl}
\usepackage{beramono}
\usepackage[T1]{fontenc}
\usepackage[latin9]{inputenc}
\usepackage{amssymb}
\usepackage{graphicx}
\usepackage{esint}

\makeatletter

\providecommand{\tabularnewline}{\\}

\makeatother

\usepackage{babel}
\begin{document}

\title{Low temperature saturation of phase coherence length in topological
insulators}

\author{Saurav Islam$^{1}$, Semonti Bhattacharyya$^{1,2}$, Hariharan Nhalil$^{1}$,
Mitali Banerjee$^{1,3}$, Anthony Richardella$^{4}$, Abhinav Kandala$^{4,5}$,
Diptiman Sen$^{6}$, Nitin Samarth$^{4}$, Suja Elizabeth$^{1}$,
Arindam Ghosh$^{1,7}$}

\affiliation{$^{1}$Department of Physics, Indian Institute of Science, Bangalore: $560012$.}

\address{$^{2}$School of Physics and Astronomy, Monash University, VIC $3800$,
Australia.}

\address{$^{3}$Department of Physics, Columbia University, New York, NY $10027$,
USA.}

\address{$^{4}$Department of Physics, The Pennsylvania State University,
University Park, Pennsylvania $16802-6300$, USA.}

\address{$^{5}$IBM T.J. Watson Research Center, Yorktown Heights, New York
$10598$, USA.}

\address{$^{6}$Center for High Energy Physics, Indian Institute of Science,
Bangalore: $560012$.}

\affiliation{$^{7}$Center for Nanoscience and Engineering, Indian Institute of
Science, Bangalore: $560012$.}
\email{isaurav@iisc.ac.in}

\thanks{SI and SB contributed equally}
\begin{abstract}
Implementing topological insulators as elementary units in quantum
technologies requires a comprehensive understanding of the dephasing
mechanisms governing the surface carriers in these materials, which
impose a practical limit to the applicability of these materials in
such technologies requiring phase coherent transport. To investigate
this, we have performed magneto-resistance\ (MR) and conductance
fluctuations\ (CF) measurements in both exfoliated and molecular
beam epitaxy grown samples. The phase breaking length\ ($l_{\phi}$)
obtained from MR shows a saturation below sample dependent characteristic
temperatures, consistent with that obtained from CF measurements.
We have systematically eliminated several factors that may lead to
such behavior of $l_{\phi}$ in the context of TIs, such as finite
size effect, thermalization, spin-orbit coupling length, spin-flip
scattering, and surface-bulk coupling. Our work indicates the need
to identify an alternative source of dephasing that dominates at low
$T$ in topological insulators, causing saturation in the phase breaking
length and time. 
\end{abstract}
\maketitle
\begin{figure}
\includegraphics[scale=1.1]{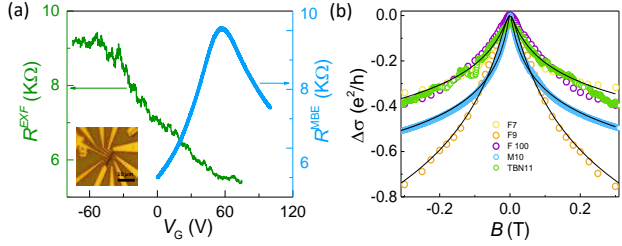}

\caption{\textbf{Quantum transport in topological insulator FETs.}\ (a) Typical
$R$\ -\ $V_{G}$ for exfoliated TI\ ($R^{EXF}$) TBN$11$ and
epitaxially grown TI\ ($R^{MBE}$) M$10$ at $20$mK. Inset:\ optical
micrograph of a typical exfoliated TI FET (b)\ Weak-anti-localisation
behavior observed in different samples at $T=300$\ mK. The solid
black lines are fits to the data using Eq.\ \ref{eq:HLN}.\label{fig:Quantum transport}}
 
\end{figure}

Topological insulators\ (TIs)\ \cite{hasan2010colloquium,moore2010birth,konig2007quantum,chen2009experimental}
are a new class of materials characterized by the presence of gapless
and linearly dispersing metallic surface states present in the bulk
band gap due to non-trivial topology of the bulk band structure.
The surface carriers are prohibited from back-scattering against non-magnetic
impurities and exhibit a plethora of fundamentally important effects
such as spin-momentum locking, hosting Majorana fermions in the presence
of a superconductor, topological magnetoelectric effect, and quantum
anomalous Hall effect\ \cite{chang2013experimental,hasan2010colloquium}.
The topological protection of these surface states makes these materials
a strong contender for the building blocks of qubits, which require
long phase coherence length\ ($l_{\phi}$) for error tolerant quantum
computation. Hence, it is critical to understand the mechanisms responsible
for dephasing or decoherence, which is equivalent to loss of information,
in the surface states of TIs. The most common dephasing mechanism
in TIs at low temperature\ ($T$) has been known to be electron-electron
interaction\ \cite{kim2011thickness,ockelmann2015phase,wang2011evidence,zhang2012interplay,kandala2013surface},
and the coupling of the surface states to localized charged puddles
in the bulk\ \cite{liao2017enhanced}. Li et al. have demonstrated
that electron-phonon interaction is also required to explain the dependence
of $l_{\phi}$ on $T$\ \cite{li2012two}. Although theoretically,
all these mechanisms lead to a diverging $l_{\phi}$ with decreasing
$T$\ \cite{checkelsky2011bulk,lin2002recent,kim2011thickness,ockelmann2015phase},
experimentally, the increase of $l_{\phi}$ with reducing $T$ is
often followed by its saturation for $T$ $\leq2-5$~K\ \cite{liao2017enhanced,steinberg2011electrically,li2012two,islam2018universal}.
The saturation of $l_{\phi}$ at a finite value instead of its divergence
for $T\rightarrow0$\ K, which is predicted for electron-electron
or electron-phonon interactions, has been a matter of active discourse\ \cite{fukai1990saturation,lin1987electron,vranken1988enhanced,fournier2000anomalous,pivin1999saturation,schopfer2003anomalous,mohanty1997intrinsic,pierre2002dephasing,pierre2003dephasing,mohanty1997decoherence,lin2002recent,chuang2013mesoscopic,huard2005effect}.

\begin{table*}
\caption{\textbf{Details of measured devices}: The thickness, substrate, composition,
saturation value of temperature\ ($T^{sat}$), and saturation value
of phase breaking length\ ($l_{\phi}^{sat}$) for various devices
that been investigated is provided in the table. }

\begin{tabular}{|c|c|c|c|c|c|}
\hline 
Sample & Thickness & Substrate & Composition & $T^{sat}$ & $l_{\phi}^{sat}$ \tabularnewline
\hline 
\hline 
F$7$ & $7$ & SiO$_{2}$/Si$++$ & Bi$_{1.6}$Sb$_{.4}$Te$_{2}$Se & $\sim2$\ K & $200$\ nm\tabularnewline
\hline 
F$100$ & $100$ & SiO$_{2}$/Si$++$ & Bi$_{1.6}$Sb$_{.4}$Te$_{2}$Se & $\sim2$\ K & $150$\ nm\tabularnewline
\hline 
F$9$ & $9$ & SiO$_{2}$/Si$++$ & Bi$_{1.6}$Sb$_{.4}$Te$_{2}$Se & $\sim2$\ K & NA\tabularnewline
\hline 
TBN$11$ & $11$ & Boron nitride & Bi$_{1.6}$Sb$_{.4}$Te$_{2}$Se & $\sim2$\ K & $190$\ nm\tabularnewline
\hline 
M$10$ & $10$ & STO & (Bi,Sb)$_{2}$Te$_{3}$ & $\sim300$\ mK & $2000$\ nm\tabularnewline
\hline 
\end{tabular}
\end{table*}

\begin{figure}
\includegraphics[scale=1.1]{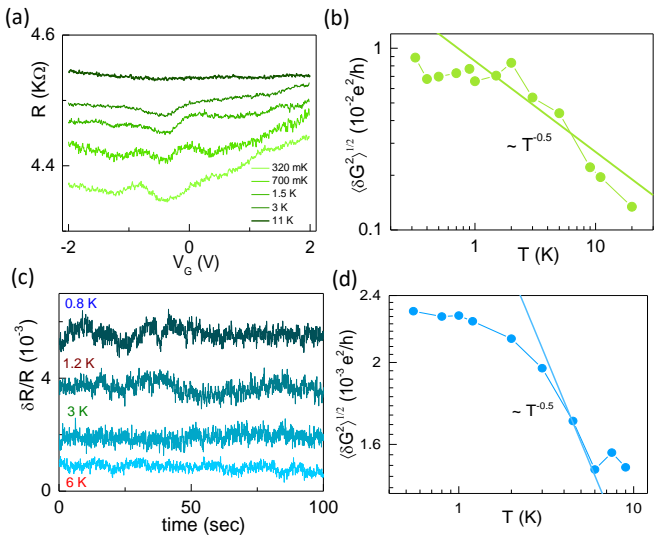}\caption{\textbf{Conductance fluctuations measurements.}\  (a)\ $R$
vs $V_{G}$ for different $T$ for device F100 in a $4$\ V window,
used to extract $\langle\delta G^{2}$. (b)\ $\langle\delta G^{2}\rangle^{1/2}$
as a function of $T$ which shows a saturation below $T<2$\ K for
the device F$100$. Above $2$\ K, $\langle\delta G^{2}\rangle^{1/2}\propto1/T^{0.5}$.
(c)\ Normalized resistance fluctuations in the time domain for different
$T$ used for calculation of P.S.D (d)\ Noise magnitude as a function
of $T$ also showing a saturation below $T<2$\ K.~Above $2$\ K,
$\langle\delta G^{2}\rangle^{1/2}\propto1/T^{0.5}$.\label{fig:noise}}
\end{figure}

\begin{figure*}
\includegraphics[scale=1.2]{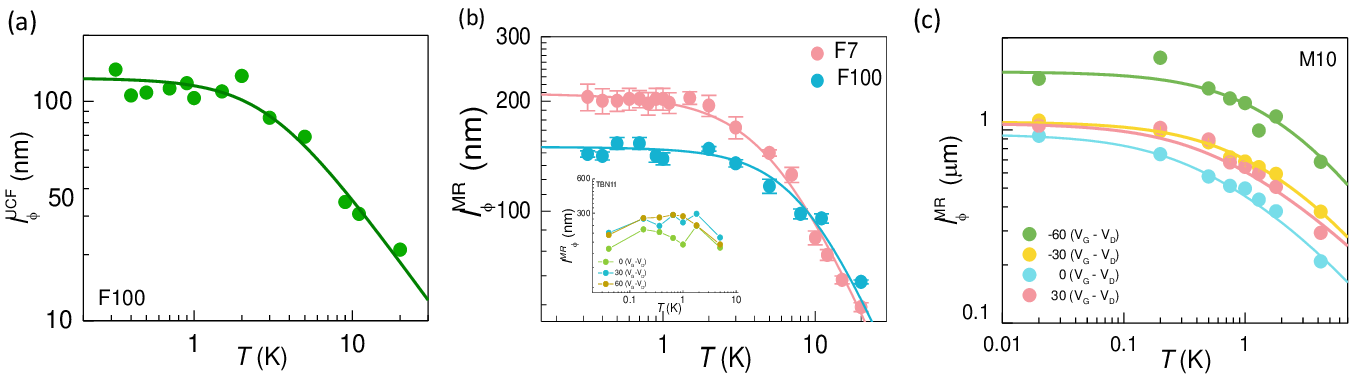}\caption{\textbf{Saturation of phase breaking length\ ($l_{\phi}$).}\ (a)\ $l_{\phi}$
vs $T$ extracted from $T$-dependence of $\langle\delta G^{2}\rangle$using
Eq.\ \ref{eq:UCF direct} for device F$100$. The solid line is fit
according to Eq.\ 4. (b)\ $l_{\phi}$ vs $T$ extracted from $T$-dependence
of WAL data using Eq.\ \ref{eq:HLN} for exfoliated samples F7 and
F100 which shows a saturation below $T<3$\ K. The solid lines are
fits to Eq.\ \ref{eq:HLN}. Inset shows \ $l_{\phi}$ vs $T$ extracted
from $T$-dependence of WAL data for sample TBN11, which also shows
a saturation. (c)\ $l_{\phi}$ vs $T$ for different $V_{G}$ extracted
from $T$-dependence of WAL data using Eq.\ \ref{eq:HLN} for epitaxially
grown TI M10 which shows a saturation below $T<300$\ mK. The solid
lines are fits to Eq.\ \ref{eq:lphi vs T}. \label{fig:Saturation-of-phase}}
\end{figure*}

In this report, we have inspected the factors that can lead to the
saturation of $l_{\phi}$ in TIs by measuring both gate-voltage\ ($V_{G}$)
and time\ ($t$)-dependent conductance fluctuations and magneto-resistance\ (MR)\ \cite{hikami1980spin,birge1990conductance,chuang2013mesoscopic,ghosh2000universal,ghosh2004set,lee1985universal,lee1987universal,pal2012direct,shamim2017dephasing}.
Conductance fluctuations result from the quantum interference of different
electron trajectories, manifested as sample specific, aperiodic fluctuations
in the conductance due to varying disorder configuration, Fermi energy,
and magnetic field; such fluctuations have been used as a tool to
probe the presence of time-reversal symmetry (TRS) breaking disorders,
since the saturation of $l_{\phi}$ at low $T$ is often attributed
to spin-flip scattering processes. The magnitude of the conductance
fluctuations~$\langle\delta G^{2}\rangle$, however, shows a factor
of two reduction upon application of a perpendicular magnetic field\ ($B_{\perp}$),
implying that TRS is intrinsically preserved in these systems. Additionally,
at different gate voltages, $\langle\delta G^{2}\rangle$ displays
a saturation for $T<2$\ K, even in the presence of a large $B_{\perp}$
which suppresses spin-spin scattering; this implies that neither magnetic
impurities nor the coupling of the surface and the bulk impurity states
is responsible for the saturation. Our experiment suggests an unconventional
mechanism that saturates $l_{\phi}$ in TIs, possibly arising from
unscreened Coulomb fluctuations from the charged disorders present
in the bulk\ \cite{liao2017enhanced}.

The field effect devices investigated in this paper were fabricated
from both exfoliated and molecular beam epitaxy\ (MBE) grown TIs.
To fabricate the former, the TI Bi$_{1.6}$Sb$_{.4}$Te$_{2}$Se~(BSTS)
(purity of the starting elements Bi, Te, Sb, Se $\geq4$N) was exfoliated
from a single crystal onto a SiO$_{2}$/Si$++$ substrate with the
$285$\ nm thick SiO$_{2}$ acting as the back gate dielectric inside a glove box\ \cite{taskin2011observation}.
This was followed by standard electron-beam lithography and sputtering
of $100$\ nm Au to form the source-drain contacts\ (inset of Fig.\ 1(a)).
The details of the devices measured are provided in Table.\ I. In
sample TBN$11$, the TI flake was transferred onto an atomically flat
boron nitride\ (BN) substrate to reduce the effect of charged traps
and dangling bonds of SiO$_{2}$ on the electrical transport\ \cite{dean2010boron},
followed by lithography and metallization. The quarternary alloy BSTS
offers a reduced bulk number density due to compensation doping, resulting
in a higher percentage of surface transport\ \cite{taskin2011observation}.
The exfoliated samples were covered with PMMA\ (poly(methylmethacrylate))
during the entire measurement cycle to prevent oxidation and subsequent
degradation of the surface quality. The large area\ ($0.5$\ mm\ $\times$\ $1$\ mm)
sample\ (M10) was fabricated from thin (thickness, $d=10$\ nm)
(Bi,Sb)$_{2}$Te$_{3}$\ (BST) (purity of starting elements was $5$N
for Bi, $5$N for Sb and $6$N for Te),) grown by molecular beam epitaxy\ (MBE)
on $\langle111\rangle$ SrTiO$_{3}$\ (STO)\ substrate and mechanically etched into a Hall bars with a metallic coating
of Indium at the back, that was used as a back gate electrode\ \cite{bhattacharyya2016resistance,islam2017bulk}.
Resistivity measurements were performed in a low-frequency four-probe
AC configuration in a pumped He-$3$ system (base $T=320$\ mK) and
in a dilution refrigerator (base$T=20$\ mK).

Preliminary electrical transport characteristics in the exfoliated
device TBN$11$\ (at $T=20$\ mK) and the MBE-grown device (at $20$\ mK)
are shown in Fig.\ 1(a). The $R$-$V_{G}$ data indicates that at
$V_{G}=0$\ V, TBN$11$ is intrinsically electron doped and M$10$
is intrinsically hole doped. Whereas M$10$ shows a clear graphene
like ambipolar transport with a Dirac point at $60$\ V, which could
be achieved due to the high dielectric constant of the STO at low
$T$\ ($\epsilon_{r}\sim44000$ at $5$\ K)\ \cite{islam2017bulk},
TBN$11$ shows a clear signature of an electron-hole puddle regime
at $V_{G}=-60$\ V. The estimated value of intrinsic number density
at $V_{G}=0$\ V are $-2.9\times10^{13}$\ m$^{-2}$ and $9\times10^{13}$\ m$^{-2}$
respectively. The quantitative difference of the $V_{G}$-dependent
characteristics here indicates dominance of different types of disorder
species in the samples, owing to different processes of synthesis,
fabrication, and composition. The n-doping in BSTS mostly comes due
to Se vacancies, the most likely probable cause of p-doping in the
epitaxially grown samples is the presence of anti\textendash site
defects.

Fig.\ 1(b) shows weak anti-localization\ (WAL) for different samples,
which is characterized by a cusp in the quantum correction to conductivity
$\triangle\sigma$ at $B=0$\ T, and is a result of $\pi$ Berry
phase in topological insulators. The magneto-conductance data can
be fitted with the Hikami-Larkin-Nagaoka\ (HLN) expression for diffusive
metals with high spin-orbit coupling\ $(\tau_{\phi}>>\tau_{so},\tau_{e})$\ \cite{hikami1980spin,bao2012weak}:

\begin{equation}
\triangle\sigma=-\alpha\frac{e^{2}}{\pi h}\left[\psi\left(\frac{1}{2}+\frac{B_{\phi}}{B}\right)-\ln\left(\frac{B_{\phi}}{B}\right)\right]\label{eq:HLN}
\end{equation}
where $\tau_{\phi}$, $\tau_{so}$, $\tau_{e}$ are the phase coherence
or dephasing time, spin-orbit scattering time and elastic scattering
time respectively, $\psi$ is the digamma function and $B_{\phi}$
is the phase breaking field. Here $\alpha$ and $B_{\phi}$ are fitting
parameters. The phase coherence length $l_{\phi}^{MR}$ can be extracted
using $l_{\phi}^{MR}=\sqrt{\hbar/4eB_{\phi}}$ and the value of $\alpha$
gives an estimate of the number of independent conducting channels
in the sample\ (See supplementary information).

The magnitude of gate voltage dependent conductance fluctuations\ $\langle\delta G^{2}\rangle$,
has been evaluated by using a method similar to Ref.\ \cite{gorbachev2007weak,pal2012direct,islam2018universal}
by varying the chemical potential with the back gate voltage in steps
of $5$\ mV over a small window of $4$\ V. $\langle\delta R^{2}\rangle$
is extracted from $R$\ -\ $V_{G}$ by fitting the data with a smooth
polynomial curve. $\langle\delta G^{2}\rangle$ is then obtained using
the relation:$\langle\delta G^{2}\rangle=\langle\delta R^{2}\rangle/\langle R\rangle^{4}$\ ($\langle\delta R^{2}\rangle$
is extracted from the variance of the residual). As shown in Fig.~\ref{fig:noise}(a)
for a typical $4$\ V window, the fluctuations are aperiodic yet
reproducible but weaken with increasing $T$. The $T$-dependence
of the standard deviation $\left\langle \delta G^{2}\right\rangle ^{\frac{1}{2}}$,
at $V_{G}=0$~V (center of the corresponding window) for F$100$
in Fig.~2(b) shows two distinctly different regions. Above $T>2$\ K,
$\left\langle \delta G^{2}\right\rangle ^{\frac{1}{2}}\propto T^{-0.5}$,
which is expected from the $T$-dependence of $l_{\phi}\propto T^{-0.5}$
and the number of active scatterers\ ($n_{s}\propto T$)\ \cite{islam2018universal}.
$\left\langle \delta G^{2}\right\rangle ^{\frac{1}{2}}$, however,
saturates for $T<2$\ K.

Conductance fluctuations due to changes in the disorder configuration
were detected by measuring the normalized noise magnitude, defined
as $\frac{\left\langle \delta G^{2}\right\rangle }{\langle G^{2}\rangle}=\frac{\int S_{V}df}{V^{2}}$
where $S_{V}/V^{2}$ is the normalized power spectral density\ (P.S.D.)
of the time-dependent signal, in an AC-four probe Wheatstone bridge
technique\ (\cite{scofield1987ac,ghosh2004set}). The normalized
time-dependent fluctuations in resistance, ($\delta R/R$) for various
temperatures ($0.3$~K\ $\leq T\leq6$~K) is shown in Fig.~\ref{fig:noise}(c).
$\left\langle \delta G^{2}\right\rangle ^{\frac{1}{2}}$ extracted
from the P.S.D. of time-dependent conductance fluctuations is plotted
as a function of $T$ in Fig.~\ref{fig:noise}(d) and is found to
show a saturation below $T=2$~K, which is consistent with the behavior
of $\left\langle \delta G^{2}\right\rangle ^{\frac{1}{2}}$ obtained
from the $V_{G}$-dependence. The order of magnitude difference is
caused due to integration of the signal over a finite frequency window
as well as the sensitivity of resistance changes to individual defect
movements\ \cite{shamim2017dephasing,birge1989electron,trionfi2004electronic}.

The phase breaking length, $l_{\phi}$ extracted from $\langle\delta G^{2}\rangle^{\frac{1}{2}}$-$T$
data\ (Fig.~\ref{fig:noise}(a)) using the expression\ \cite{akkermans2007mesoscopic,adroguer2012diffusion}
\begin{equation}
\langle\delta G^{2}\rangle\simeq\left(\frac{3}{\pi}\right)\left(\frac{e^{2}}{h}\right)^{2}\left(\frac{l_{\phi}}{L}\right)^{2}\label{eq:UCF direct}
\end{equation}
is shown in Fig.\ \ref{eq:lphi vs T}a. Since $\langle\delta G^{2}\rangle\propto l_{\phi}^{2}$,
any saturation obtained from time or gate voltage-dependence should
also be reflected in the saturation of $l_{\phi}$, obtained directly
from MR measurements. The values of $l_{\phi}$ extracted from MR
data similar to Fig.\ \ref{fig:Quantum transport}(b), as a function
of $T$ for the exfoliated samples F$100$ and F$7$ is shown in Fig.~3(b).
We find that $l_{\phi}$ obtained from two different methods, $l_{\phi}^{MR}$
and $l_{\phi}^{UCF}$\ (extracted from Eq.\ \ref{eq:HLN} and Eq.\ \ref{eq:UCF direct}
respectively) show similar trends with $T$, first increasing with
decreasing $T$, followed by a saturation below $T\sim2$~K, thus
discarding the possibility of the saturation to be an artifact. The
discrepancies in the values of $l_{\phi}^{MR}$ and $l_{\phi}^{UCF}$
are within uncertainties of the prefactor of Eq.\ \ref{eq:UCF direct}\ \cite{akkermans2007mesoscopic,adroguer2012diffusion,beenakker1991quantum}.
The higher value of $l_{\phi}$ for F$7$ compared to $F100$ can
be due to enhanced dephasing due to trapping-detrapping processes
in the bulk, since the thickness of F$100$ is much larger than that
of F$7$.

For a quantitative understanding, $l_{\phi}$-$T$ data\ (Fig.\ \ref{eq:lphi vs T}
(a-b)) has been fitted with the expression commonly used to fit the
$l_{\phi}$-$T$ data in 2D diffusive systems\ \cite{lin2002recent,li2012two}.
\begin{equation}
l_{\phi}=\frac{1}{(A_{0}+A_{1}T+A_{2}T^{2})^{0.5}}\label{eq:lphi vs T}
\end{equation}

\noindent Here, $l_{\phi}\propto T^{-0.5},$ and $l_{\phi}\propto T^{-1}$
are the respective contributions from electron-electron\ (e-e) and
electron-phonon\ (e-ph) scattering. $A_{1}$, $A_{2}$ are fitting
parameters and $A_{0}=\frac{1}{l_{\phi0}^{2}}$, the saturation value
of $l_{\phi}$. In 2D systems, although e-e interactions are the dominant
source of dephasing at low $T$ and have been adequate to describe
dephasing in graphene\ \cite{morozov2006strong,wu2007weak,lin2002recent},
and in some reports of TI\ \cite{checkelsky2011bulk,liu2011eeinteraction},
e-ph interaction cannot be neglected for TI because of the vicinity
of the bulk to the surface states\ \cite{li2012two,liao2017enhanced}.
However, instead of saturation, these two mechanisms lead to a diverging
$l_{\phi}$ at low $T$. The saturation of $l_{\phi}$ has also been
obtained in device TBN$11$, where the TI has been transferred onto
a boron nitride substrate\ (Fig.\ \ref{eq:lphi vs T}b). The atomically
flat boron nitride\ (thickness $d=14$\ nm) prevents trapping-detrapping
that is commonly observed between the channel and the SiO$_{2}$ substrate,
thus reducing any dephasing due to electromagnetic fluctuations induced
by potential traps present in SiO$_{2}$.

The saturation of $l_{\phi}$ is often attributed to (a)\ finite
size effects, where $l_{\phi}$ becomes comparable to $L$, the length
of the sample, (b)\ saturation of electron temperature due to heating
from external sources\ \cite{lin2002recent}, (c)\ spin-orbit scattering
length becoming comparable to the phase breaking length\ \cite{fukai1990saturation},
and (d)\ magnetic impurities or local magnetic moment\ \cite{pierre2002dephasing,pierre2003dephasing,schopfer2003anomalous}.
We have systematically explored the possibility of saturation arising
from any of the above reasons. To probe the effect of finite size,
we have performed MR measurements on M$10$ with a channel length
of $1$\ mm which is three orders of magnitude more than the saturated
value of $l_{\phi}$. The $l_{\phi}$ extracted from magneto-conductance
data using Eq.\ \ref{eq:HLN} also shows a similar saturation for
$T<300$\ mK for all $V_{G}$'s. The saturation obtained in the large
area sample, indicates that finite-size effects at low $T$ is not
the cause of the observed behavior of $l_{\phi}$. Another important
factor for the saturation is thermalization, where the electron $T_{e}$
can be much higher than the lattice temperature $T_{L}$. For extracting
$T_{e}$ accurately in our He-$3$ system, we have measured and analyzed
the $T$ dependence of Shubnikov-de Haas oscillations using GaAs/Al$_{0:33}$Ga$_{0:66}$As
hetero-structure to ensure that down to $T=0.3$\ K, $T_{e}$ matches
$T_{L}$. This type of saturation has also been observed in systems
where the spin-orbit coupling length\ ($l_{SO}$) becomes comparable
to the phase breaking length\ \cite{fukai1990saturation}. We have
extracted $l_{SO}$ by using the full HLN equation\ \cite{zhang2012magneto,dey2014strong}\ (See
supplementary information). The extracted $l_{SO}$ is much smaller
compared to $l_{\phi}$ (as expected for TI systems) ruling out that
possibility as well.

\begin{figure}
\includegraphics[scale=1.2]{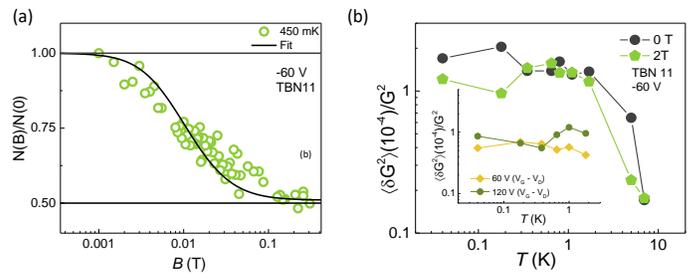}\caption{\textbf{Variation of UCF with magnetic field.}\ (a)\ Normalized
UCF magnitude at $V_{G}-V_{D}=0$\ V for $450$\ mK clearly exhibiting
a factor of two reduction indicating the intrinsic preservation of
time reversal symmetry in these systems. The solid line is fit to
the data according to Eq.\ \ref{eq: UCF Fitting}. (b)\ Normalized
variance $\langle\delta G^{2}\rangle/\langle G^{2}\rangle$ as function
of $T$ for $B_{\perp}=0$\ T and $2$~T at $V_{G}-V_{D}=-60$\ V.
The qualitative nature of saturation does not change even upon the
application of a high magnetic field, probably indicating that magnetic
moments are not responsible for the saturation. Inset shows normalized
variance at $V_{G}-V_{D}=60$\ V and at $120$\ V, both showing
similar saturation as a function of $T$.\label{fig:Variation-of-UCF}}
\end{figure}

The most common reason, however, for the saturation of $l_{\phi}$
is the presence of magnetic impurities or localized spins, which leads
to the saturation even in extremely pure systems\ \cite{pierre2002dephasing,pierre2003dephasing,schopfer2003anomalous}.
Experimental and theoretical studies have reported the presence of
localized spins\ \cite{nisson2013nuclear} or intrinsic magnetic
instabilities in the TI surface\ \cite{baum2012magnetic}. One manifestation
of the presence of magnetic impurities can be long- or short-range
magnetic ordering , which can lead to the removal of TRS in the system.
To probe this, we have measured $\langle\delta G^{2}\rangle$ as a
function of $B_{\perp}$ on TBN$11$. The magnitude of the conductance
fluctuations is plotted as $N_{G}(B)/N_{G}(0)$\ ($N_{G}=\langle\delta G^{2}\rangle/\langle G^{2}\rangle$
is the normalized variance) in Fig.\ \ref{fig:Variation-of-UCF}(a).
We observe a factor of two reduction in the normalized magnitude when
$B_{\perp}>>B_{\phi}$ due to the suppression of the Cooperon contribution
in transport and the crossover of the system from symplectic to unitary
symmetry class, which indicates that TRS is intact intrinsically in
these systems. For a quantitative understanding, the normalized magnitude
has been fitted with the expression\ \cite{stone1989reduction,shamim2017dephasing,lee1987universal,islam2018universal}

\[
N(B)/N(0)=\frac{1}{2}+\frac{1}{b^{2}}\sum_{n=0}^{\infty}\frac{1}{[\left(n+\frac{1}{2}\right)+\frac{1}{b}]^{3}}
\]
\begin{equation}
=\frac{1}{2}-\frac{1}{2b^{2}}\Psi^{''}\left(\frac{1}{2}+\frac{1}{b}\right).\label{eq: UCF Fitting}
\end{equation}
Here $b=8\pi B(l_{\phi})^{2}/(h/e)$, $\Psi^{''}$ is the double derivative
of the digamma function, and $l_{\phi}$ is the fitting parameter.
The solid line in Fig.\ 4 (a) is the fit according to Eq.\ \ref{eq: UCF Fitting},
which captures the variation with normalized magnitude with $B$.
The $l_{\phi}$ obtained from the fit is $220$\ nm, similar to $l_{\phi}$
obtained from UCF and MR, which confirms the validity of the analysis
and the factor of two reduction.

Spin-flip scattering due to the presence of unwanted magnetic impurities,
e.g., due to finite purity of the metal components, may also cause
dephasing. To investigate this, we have extracted $\langle\delta G^{2}\rangle$
as a function of $T$ for $B_{\perp}=0$\ T and $2$\ T at different
$V_{G}-V_{D}$. The normalized magnitude $\langle\delta G^{2}\rangle/G^{2}$
shows a saturation for both $B_{\perp}=0$\ T and $2$\ T for $T<2$\ K.
The presence of a magnetic field $B_{\perp}>>k_{B}T/(g\mu_{B})$ is
expected to freeze the magnetic moments and suppress spin-flip scattering
in the sample. Here $k_{B}$, $g$, and $\mu_{B}$ are the Boltzmann
constant, Land$\acute{e}$ g-factor, and Bohr magneton respectively. The fact
that the saturation persists even in the presence of $B_{\perp}$,
indicates that it is not due to any magnetic impurities or localized
spins in the system. To probe the effect of coupling between the surface
states and charged disorders in the bulk, we have measured the conductance
fluctuations for more positive gate voltages\ ($V_{G}-V_{D}=60$\ V
and $120$\ V). The number density at $V_{G}-V_{D}=120$\ V is $8.6\times10^{16}$\ m$^{-2}$,
which corresponds to bulk transport dominated regime, where the coupling
of the two surfaces and the bulk is also higher, compared to that
near the Dirac point\ ($-60$\ V)\ \cite{kim2012surface,liao2017enhanced}.
However, the nature of saturation also does not change for more positive
gate voltages, implying that it is independent of the coupling between
surface states and the charged puddles in the bulk\ (inset of Fig.\ \ref{eq: UCF Fitting}b).

While the exact source of saturation remains unascertained, we discuss
some plausible mechanisms that may lead to the saturation of $l_{\phi}$
in TIs. The saturation can arise from the presence of two-level systems\ as
has been explored in Ref.\ \cite{imry1999low,aleshin2001low}. Such
two-level systems can arise from the charge fluctuations in the bulk,
which are known to be the dominant source of $1/f$ noise in TI\ \cite{bhattacharyya2015bulk,islam2017bulk,bhattacharyya2016resistance}.
The relaxation dynamics of these charged defects in the bulk can lead
to a very weak dependence of $l_{\phi}$ on $T$. We also note that
the temperature where $l_{\phi}$ saturates in M$10$ is $\sim300$\ mK,
which is an order of magnitude lower compared to the exfoliated samples\ (F$7$
and F$10$), and $l_{\phi}^{sat}$ is also an order of magnitude higher.
This difference in $l_{\phi}^{sat}$ and $T^{sat}$ between these
samples, grown by totally different methods could be indicative of
a lower charge impurity driven inhomogeneity in the bulk in case of
the MBE sample. Liao et al. have shown that the charge puddles in
the bulk lead to a sublinear dependence of $l_{\phi}$ on $T$\ \cite{liao2017enhanced}.
It is possible that at low $T$, these uncompensated charges are strongly
localized, leading to reduced screening of electromagnetic fluctuations.
This can produce additional dephasing of the surface carriers, which
might limit $l_{\phi}$ to a finite value. Recently V\"{a}yrynen et al.
have proposed back-scattering of electrons by electromagnetic fluctuations
from the charge puddles in the bulk in 2D TIs\ \cite{vayrynen2018noise}.
The effect of such inelastic scattering on $l_{\phi}$ in 3D TIs remains
to be seen, and may also provide crucial insight into the factors
leading to the saturation of $l_{\phi}$ in topological insulators.

In conclusion, we have measured the Fermi energy and time-dependent
conductance fluctuations and magneto-resistance to probe the sources
of dephasing of the surface carriers in topological insulators. The
phase breaking length obtained from both these techniques show a saturation
below some specific temperatures which are sample dependent. We have
eliminated several factors that may lead to the saturation of $l_{\phi}$
such as finite-size effects, spin-orbit coupling length, and surface-bulk
coupling. The magnetic field dependence of the conductance fluctuations
also eliminates the possibility of the saturation arising due to the
presence of magnetic impurities or localized spins in the system.
Our work suggests an additional dephasing mechanism in TIs which is
dominant at low temperatures, and limits the phase breaking length
to a finite value at low temperatures.

S.I., S.B., D.S., and A.G. acknowledge support from DST, India. A.R.,
A.K., and N.S. acknowledge support from The Pennsylvania State University
Two-Dimensional Crystal Consortium \textendash{} Materials Innovation
Platform (2DCC-MIP), which is supported by NSF cooperative Agreement
No. DMR-1539916.

\bibliographystyle{apsrev4-1}

\end{document}